\documentstyle[prl,aps,epsf]{revtex}
\begin{document}
\input epsf
\draft
\twocolumn[\hsize\textwidth\columnwidth\hsize\csname@twocolumnfalse\endcsname

\title {First Order Transition in the Spin Dynamics of Geometrically
Frustrated Yb$_2$Ti$_2$O$_7$.}

\author {J.A. Hodges$^1$, P. Bonville$^1$, A. Forget$^1$, 
A. Yaouanc$^2$, P. Dalmas de R\'eotier$^2$, G. Andr\'e$^3$, \\
M. Rams$^4$, K. Kr\'olas$^4$, C. Ritter$^5$,
P.C.M. Gubbens$^6$, C.T. Kaiser$^6$, P.J.C. King$^7$, C. Baines$^8$}
\address{$^1$CEA - Saclay, DRECAM - SPEC, 91191 Gif sur Yvette, France}
\address{$^2$CEA - Grenoble, DRFMC - SPSMS, 38054 Grenoble, France}
\address{$^3$Laboratoire L\'eon Brillouin, CEA - CNRS, 91191 Gif sur Yvette, 
France}
\address{$^4$Institute of Physics, Jagiellonian University, Krak\'ow, Poland}
\address{$^5$Institute Laue-Langevin, Grenoble, France}
\address{$^6$Interfacultair Reactor Instituut, TU - Delft, 2629 JB Delft, 
The Netherlands}
\address{$^7$Rutherford Appleton Laboratory, Chilton, OX11 0QX, United Kingdom}
\address{$^8$Laboratory for Muon Spectroscopy, Paul Scherrer Institute, 5232 Villigen-PSI, Switzerland}
\date{\today} \maketitle 

\begin{abstract}
Using neutron diffraction, $^{170}$Yb M\"ossbauer and $\mu$SR 
spectroscopies, we have examined the pyrochlore Yb$_2$Ti$_2$O$_7$ where the
Yb$^{3+}$ S$'$ = 1/2 ground state has planar anisotropy. Below $\sim$ 0.24 K,
the temperature of the known specific heat $\lambda$-transition, there is no 
long range magnetic order. 
We show that the transition corresponds to a first
order change in the fluctuation rate of the Yb$^{3+}$ spins. Above the 
transition temperature, the rate, in the GHz range, follows a thermal 
excitation law, whereas below, the rate, in the MHz range, is temperature 
independent indicative of a quantum fluctuation regime.

\end{abstract}


]

Geometrically derived magnetic frustration arises when the spatial 
arrangement of the spins is such that it prevents the simultaneous 
minimisation of all interaction energies 
\cite{diep,schiffer96,moessner,villain}. 
In the crystallographically ordered pyrochlore structure compounds 
R$_2$Ti$_2$O$_7$, the rare earth ions (R) form a sub-lattice of
corner sharing tetrahedra and a number of these compounds 
have been observed to exhibit frustration related behaviour 
\cite{raju,harris1,harris2,gardner,gingras,denhertog,ramirez,bramwellCM}.
The low temperature magnetic behaviour associated with a particular rare 
earth depends on the properties of the crystal field ground state and on 
the origin (exchange/dipole), size and sign of the inter-ionic interactions.
For example, the recently identified spin-ice configuration
\cite{harris1,denhertog,ramirez} has been linked with an Ising-like 
anisotropy and a net ferromagnetic interaction.
Most of the published work on the pyrochlores has concerned rare earth ions 
with Ising-like characteristics 
\cite{harris1,harris2,gardner,gingras,denhertog,ramirez,bramwellCM} 
and there has also been some interest in the properties of
weakly anisotropic Gd$^{3+}$ \cite{raju}. Less attention has been paid to 
the case, considered here, where the ion has planar anisotropy. 

To date, in systems where geometrical frustration may be present,
two different low temperature magnetic ground states have been 
considered. First, under the influence of the frustration the system does not 
experience a magnetic phase transition and remains in a collective 
paramagnetic state with the spin fluctuations persisting as T $\to$ 0
\cite{moessner,villain,harris1,harris2,gardner,gingras,denhertog,ramirez,dunsiger}.
Second, a long range ordered state is reached through a phase transition 
which may be first order 
\cite{raju,bramwell94,melko01}. 
Our results for Yb$_2$Ti$_2$O$_7$, obtained using neutron 
diffraction, $^{170}$Yb M\"ossbauer spectroscopy and muon spin relaxation 
($\mu$SR), evidence a novel scenario: 
there is a first order transition which does not correspond to a 
transition from a paramagnetic state to a (long or short range)
magnetically ordered state. The transition chiefly concerns the time domain, 
and involves an abrupt slowing down of the dynamics of short range 
correlated spins; below the transition temperature, these spins continue to 
fluctuate at a temperature independent rate.

\begin{figure}
\epsfxsize=7cm
\centerline{\epsfbox{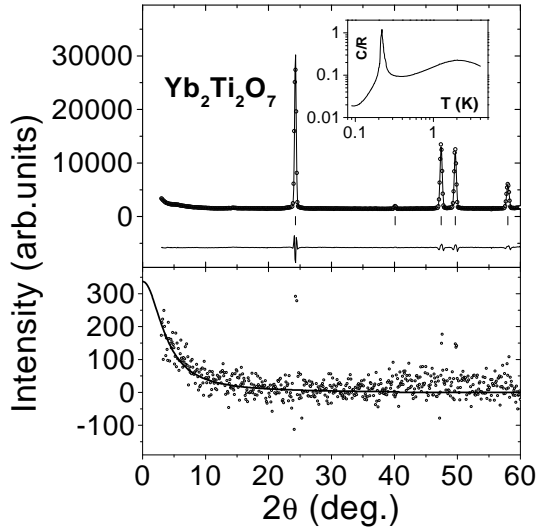}}
\caption{Neutron diffraction measurements for Yb$_2$Ti$_2$O$_7$. Top:
measured (points) and calculated (solid line) for the paramagnetic state at 
7 K; the ticks indicate the positions of the nuclear Bragg peaks and the 
difference between the measurement and the Rietveld refinement is also shown.
Bottom: the points correspond to the difference between the measured values
at 0.11 K (below the specific heat $\lambda$-transition) and at 7 K. No 
magnetic Bragg peaks are visible. The upturn observed below 
2$\theta \sim 15^{\circ}$ and the fitted solid line are discussed in the text.
The inset in the top part shows the specific heat data
taken from ref. \protect\cite{blote}.}
\label{figneutron}
\end{figure}

We have established the background magnetic characteristics 
for Yb$_2$Ti$_2$O$_7$ in a separate study \cite{hodgescef,hodgescanada}. 
The Yb$^{3+}$ ion crystal field ground state is a very well isolated Kramers 
doublet with a planar anisotropy, $g_{\perp}/g_z$ $\simeq$ 2.5 where
$g_z$ and $g_{\perp}$ are respectively the spectroscopic factors along and 
perpendicular to a local [111] axis. The net inter-ionic interaction is 
ferromagnetic (the paramagnetic Curie-Weiss temperature is 0.75\,K)
and it is dominated by exchange (the dipole-dipole interaction is relatively 
small due to the modest value, 1.15\,$\mu_B$, see below, of the Yb$^{3+}$ 
moment). Specific heat measurements \cite{blote} 
(Fig. \ref{figneutron}, top, inset) have evidenced a sharp 
peak ($\lambda$ transition) at $\sim$ 0.2\ K with an associated magnetic 
entropy of $\sim 0.18 R \ln 2$, and a broad peak centered near 2\ K. The total 
magnetic entropy was estimated to be $0.97 R \ln 2$ \cite{blote} suggesting 
there may be a small amount of missing low temperature magnetic entropy.

Single phase, polycrystalline Yb$_2$Ti$_2$O$_7$ was prepared by heating the
constituent oxides up to 1400 $^\circ$C with four intermediate grindings.
The neutron diffraction measurements were made down to 0.11\ K on the G41 line
(wavelength: 0.2427\,nm) at the Laboratoire L\'eon Brillouin 
(Fig. \ref{figneutron}) and to 0.065\ K on the D1B line at the Institute 
Laue-Langevin. Fig. \ref{figneutron} shows that the crystal structure does not
change with temperature and that below 0.2\ K, there are no magnetic Bragg 
peaks, neither isolated nor superposed on the nuclear peaks.
Had the moments of 1.15 $\mu_B$ (to be evidenced below) undergone long range 
ordering, the peaks would have easily been seen. We estimate that a magnetic 
Bragg peak would have remained visible if the correlation length had exceeded 
2 to 3\ nm, so this provides an upper bound for the 
correlation length. An additional low angle contribution
appears in the difference spectra below $\sim$ 2\ K 
(Fig. \ref{figneutron}, bottom). Its
intensity initially grows as the temperature decreases but there is no
significant change on crossing the temperature of the specific heat 
$\lambda$-transition. If we interpret this additional contribution in terms of 
short range ferromagnetic correlations, then using a model of non-interacting
spherical clusters each having a uniform magnetisation 
(solid line on Fig.~\ref{figneutron}, bottom) \cite{guinier} we obtain a 
sphere diameter of 
$\sim$ 1.5(2) nm. It is possible that not all of the additional
signal seen for 2$\theta < 15^{\circ}$ is of magnetic origin
\cite{broholm}. We consider our estimate of the magnetic correlation length
below the specific heat $\lambda$-transition to correspond to an upper limit.
The absence of long range order contradicts previous suppositions
\cite{raju,blote}.

\begin{figure}
\begin{minipage}[l]{0.23\textwidth}
\epsfxsize=3.5cm
\centerline{\epsfbox{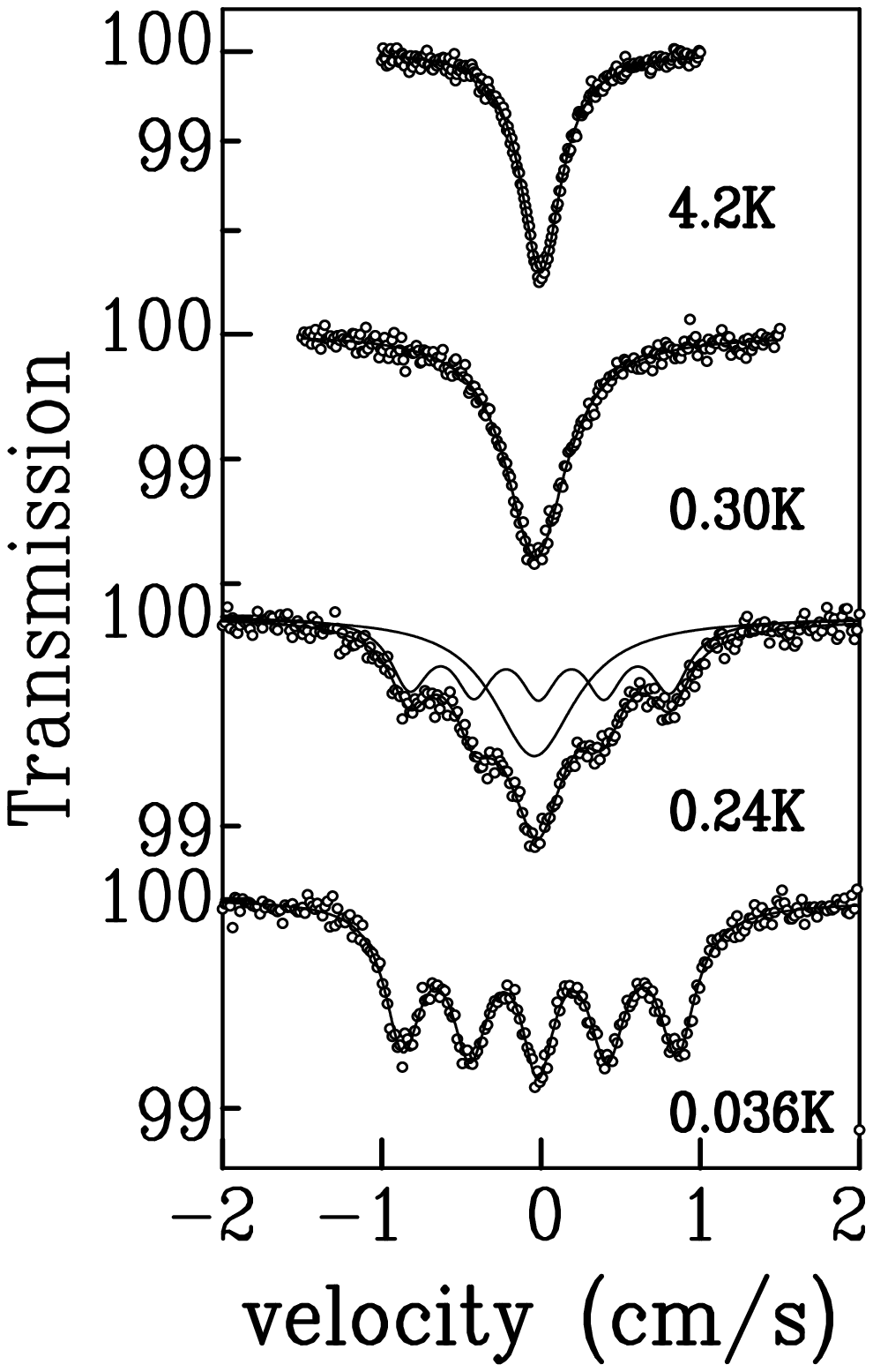}}
\end{minipage}\hfill
\begin{minipage}[r]{0.23\textwidth}
\epsfxsize=3.6cm
\centerline{\epsfbox{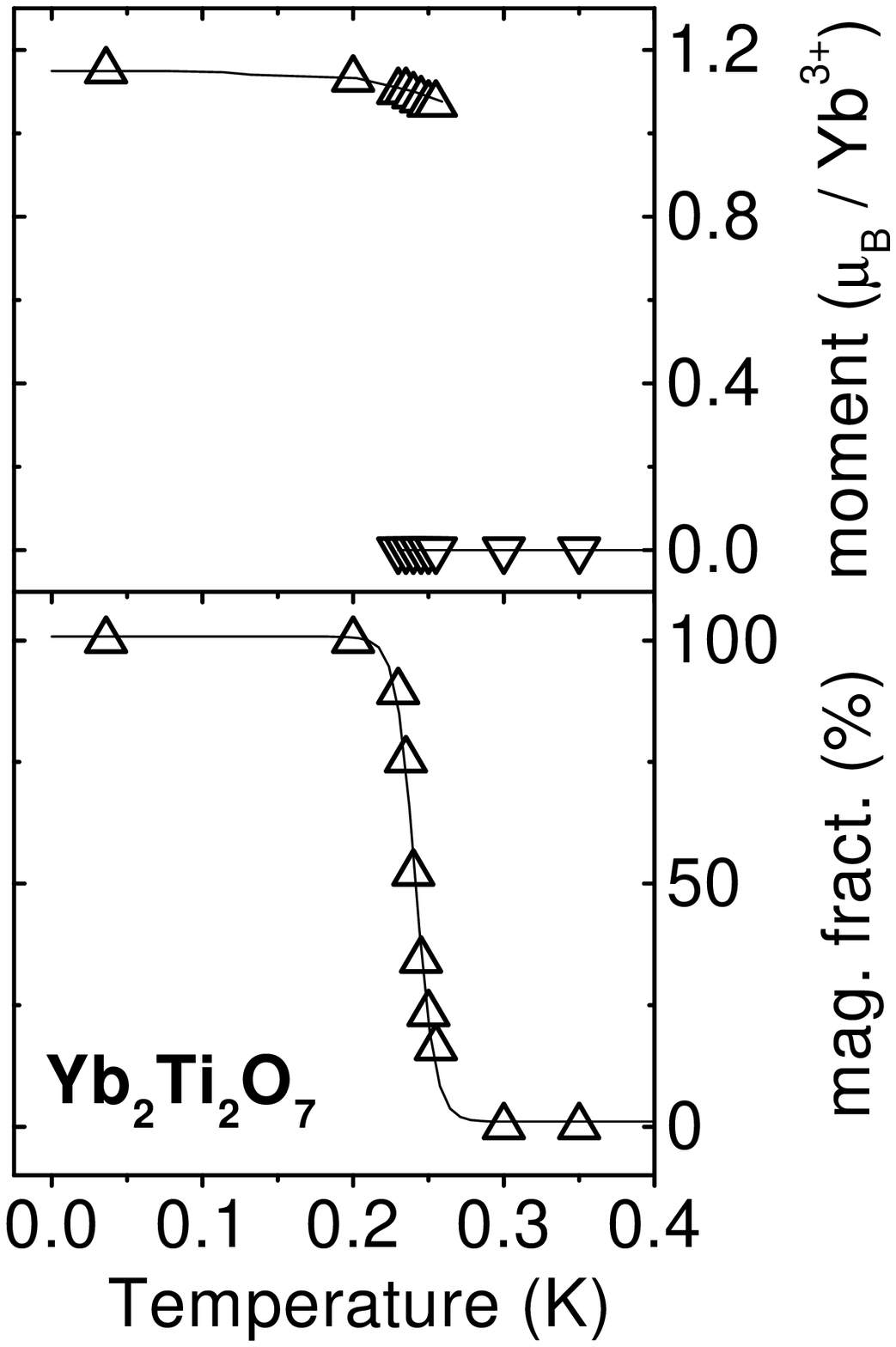}}
\end{minipage}
\caption{Left panel: $^{170}$Yb  M\"ossbauer absorption in Yb$_2$Ti$_2$O$_7$ 
above, within, and below the first order transition occurring near 0.24\ K. 
The $\gamma$-ray energy is $E_{\gamma}$ = 84\ keV, and the ground and excited 
nuclear spin states are $I_g=0$ and $I_{\rm ex}=2$, respectively (1\ cm/s 
corresponds to 680\ MHz); Right panel: thermal variation of the size of the 
Yb$^{3+}$ magnetic moment obtained from the hyperfine field (top) and 
relative weight of the ``static'' magnetic fraction (bottom). The lines are 
eye-guides.}
\label{figybmoss}
\end{figure}

Selected $^{170}$Yb M\"ossbauer absorption spectra are shown on the 
left panel of Fig. \ref{figybmoss}.
At 0.036\ K, a five line spectrum is observed indicating there is a ``static'' 
hyperfine field ($H_{\rm hf}$) which we find amounts to 115\ T. In the present 
case, ``static'' means 
the fluctuation frequency of the field is less than the lowest measurable 
$^{170}$Yb value of $\sim$ 15\ MHz. Knowing that for Yb$^{3+}$
the hyperfine field is proportional to the 4f shell magnetic moment, we 
obtain that each of the Yb$^{3+}$ carries a magnetic moment of 
1.15\ $\mu_B$. As there is no long range order, the hyperfine field 
is associated with the short range correlated Yb$^{3+}$ moments.

In the absence of a significant quadrupole hyperfine interaction, we cannot
obtain the local direction of the Yb$^{3+}$ magnetic moment by directly
measuring the angle it makes with the principal axis of the electric field 
gradient (a [111] direction). Instead, we make use of the 
property that for an anisotropic Kramers doublet, the size of the spontaneous
magnetic moment is linked to the angle $\theta$ it makes with the local 
symmetry axis (a [111] axis) through the relation: 
$ M_{\rm Yb} = {1\over 2} g_\perp \mu_B / (\cos\theta \sqrt{r^2 + 
\tan^2\theta} )$
where $r$ is the anisotropy ratio $g_{\perp}/g_z$ 
\cite{imbert}. Using the measured moment and the known $g$ values, we find 
$\theta = 44(5)^{\circ}\, $. Thus each moment does not lie perpendicular
to its local [111] axis as would be expected if the orientation were governed 
only by the crystal field anisotropy. The uniform tilting of each moment away 
from its local [111] axis is probably related to the combined influence of 
the crystal field anisotropy and of the (presumably anisotropic) 
exchange interaction with the non-collinear neighbouring moments.

When the temperature  is increased to 0.23\ K, an additional single line 
sub-spectrum appears. 
It is linked with the fraction of the Yb$^{3+}$ whose moments 
fluctuate ``rapidly'' so that the magnetic hyperfine splitting is 
``motionally narrowed''. The two sub-spectra (see Fig. \ref{figybmoss}, left 
panel at 0.24\ K) are both present up to 0.26\ K evidencing the coexistence of 
regions with ``static'' and ``rapidly fluctuating'' moments. 
Fig. \ref{figybmoss}, right panel, shows that as the temperature 
increases, there is a progressive
decrease in the relative weight of the ``static'' hyperfine field sub-spectrum.
These features clearly evidence a first order transition, taking place at 
slightly different temperatures in the different parts of the sample. 

\begin{figure}
\epsfxsize=7cm
\centerline{\epsfbox{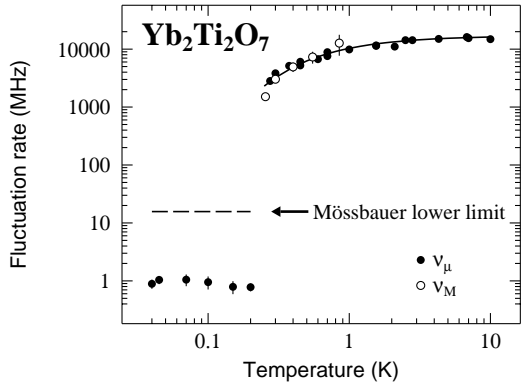}}
\caption{Estimate of the Yb$^{3+}$ fluctuation rates as
obtained from $^{170}$Yb M\"ossbauer ($\nu_M$) and $\mu$SR ($\nu_{\mu}$)
measurements. The first order change in the fluctuation rate takes place at 
the specific heat $\lambda$ transition. 
Below $\sim$ 0.24\ K, the fluctuation rate is independent of temperature and 
has dropped below the lowest value which is measurable with the M\"ossbauer 
method (dashed line). The solid line follows a thermal excitation law as 
described in the text.}
\label{figratevst}
\end{figure}

The single line (sub)spectrum progressively narrows as the temperature 
increases. As magnetic correlations are still present above 0.24 K, we 
attribute this change to the progressive increase in $\nu_M$, the fluctuation 
rate of $\vec H_{\rm hf}$. The relation between the dynamic line 
broadening, $\Delta\Gamma_R$, and $\nu_{\rm M}$, is written 
$\Delta\Gamma_R$ = $(\mu_I H_{\rm hf})^2/\nu_{\rm M}$ 
where $\mu_I$ is the $^{170}$Yb nuclear moment \cite{dattagupta}.
As shown on Fig. \ref{figratevst}, when the temperature is lowered from 
$\sim$ 1\ K to just 
above 0.24\ K, the rate decreases from $\sim$ 15 to $\sim$ 2\ GHz. 
This decrease is linked to the slowing down of the spin fluctuations which
accompanies the development of the short range correlations. 
Below 0.24\ K, $\nu_{\rm M}$ drops to a value which is less than the lowest 
measurable $^{170}$Yb M\"ossbauer value.

\begin{figure}
\epsfxsize=8.6cm
\centerline{\epsfbox{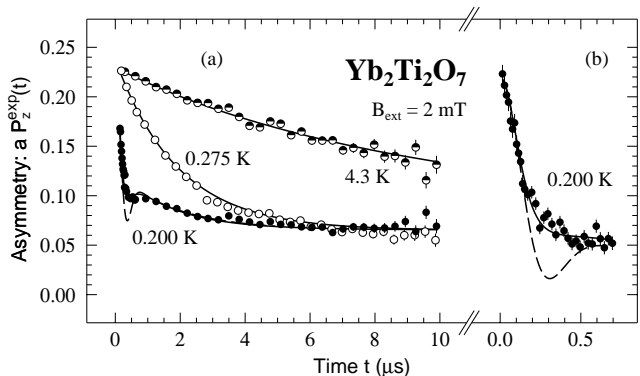}}
\caption{Panel a: $\mu$SR spectra recorded at ISIS
in a longitudinal field of 2 mT. A marked change occurs on crossing the 
temperature ($\sim$ 0.24\,K) of the specific heat $\lambda$ transition. 
Panel b: short time part of the PSI data at 0.200\,K confirming there are no 
short time oscillations. The slight difference visible between the ISIS and 
PSI data at 0.200\ K is linked with the first order nature of the transition 
and the different thermal and magnetic field histories of the two experiments.
The dashed and solid line fits are described in the text.}
\label{figmuon}
\end{figure}

The $\mu$SR study \cite{review} was carried out at the ISIS 
facility (some additional measurements were also made at the PSI facility) 
over the range 300 to 0.04\ K, mostly in a longitudinal field of 2\ mT so as 
to decouple the small contribution of the nuclear spins to the measured 
depolarisation. Fig. \ref{figmuon} 
shows typical results for the time dependence of the asymmetry 
which is written
$aP_z^{\rm exp}(t)$ = $a_zP_z(t)$ + $a_{\rm bg}$ 
where the 
first term on the right hand side originates from the sample and the second is
a temperature independent constant background contribution linked with the 
muons stopping in the silver sample holder and in the cryostat 
($a_{\rm bg} \simeq$ 0.065 for ISIS and close to 0 for PSI).

From 300 to 0.275\ K, $P_z(t)$ is well represented by an exponential 
relaxation function (solid lines at 4.3 and 0.275\ K on 
Fig. \ref{figmuon}a): $P_z(t)$ = $\exp(-\lambda_z t)$ where $\lambda_z$ is the
muon spin-lattice relaxation rate. From 100 to $\sim$ 4\ K, $\lambda_z$ is 
small ($\sim$ 0.1\ MHz) and 
in keeping with the paramagnetic nature of the Yb$^{3+}$, it shows little 
thermal dependence. As the temperature is lowered below $\sim$ 1\ K, 
$\lambda_z$ increases progressively to reach 0.52(3)\ MHz at 0.275\ K. 
The origin of this increase is the same as that for the dynamic broadening 
$\Delta\Gamma_R$ of the M\"ossbauer spectra, i.e. the slowing down of the 
electronic spin fluctuations.

As the temperature is lowered below 0.275\ K, $P_z(t)$ first becomes moderately
non-exponential and then towards 0.2\ K it abruptly becomes strongly 
non-exponential (Fig. \ref{figmuon}a). Below 0.2\ K, $P_z(t)$ is independent 
of temperature and there are no spontaneous oscillations in accordance with 
the neutron result that there is no long range order. The spectrum of 
Fig. \ref{figmuon}b which extends to shorter times, clearly confirms
this result. The shape of $P_z(t)$ below 0.2\ K is reminiscent of a dynamic 
Kubo-Toyabe (KT) decay \cite{hayano}
associated with a slowly fluctuating ensemble of moments, the decrease of 
$P_z(t)$ beyond $\sim$ 1\ $\mu$s indicating that fluctuations are 
still present. The KT decay provides an approximate account of the shape 
of $P_z(t)$ but there is a noticeable misfit below 0.5\ $\mu$s 
(dashed lines on Fig. \ref{figmuon}).
A similar misfit was also observed for Y$_2$Mo$_2$O$_7$ near 0.03\ $\mu$s at 
2.5\ K \cite{dunsiger}. A better fit is provided by the Gaussian Broadened 
Gaussian (GBG) model \cite{noakes}, where the single Gaussian distribution of 
the KT model is replaced by a collection of distributions 
(solid lines at 0.200~K on Fig. \ref{figmuon}). 
From the fit of Fig. \ref{figmuon}a, the mean value of the GBG distribution is 
$\Delta_{\rm LT}/\gamma_\mu$ $\simeq$\ 5.7 mT, the ratio of the width of the 
collection of distributions to the mean value is  $\simeq$ 0.38 and the 
electronic fluctuation rate $\nu_\mu$ is $\simeq$ 1\ MHz. All are independent 
of temperature below 0.2\ K.

To obtain an estimate of the effective value of the spin fluctuation rate, 
$\nu_\mu$, at temperatures above 0.24\ K, 
we insert the experimental values for 
$\lambda_z$ (obtained in 2\ mT) into the expression
$\lambda_z = 2\Delta_{\rm HT}^2$/$\nu_\mu$. The value for $\nu_\mu$ then 
depends on the value of $\Delta_{\rm HT}$, the Yb$^{3+}$ - muon spin 
coupling above 0.24\ K. With the choice $\Delta_{\rm HT}/\gamma_{\mu}$ = 
31.8\ mT, the $\nu_\mu$ obtained from the muon 
measurements scale remarkably well with the $\nu_{\rm M}$ 
obtained from the M\"ossbauer measurements (Fig. \ref{figratevst}). 
Their common thermal dependence is well fitted (solid line on 
Fig. \ref{figratevst}) by the thermal excitation law 
$\nu = \nu_0 \exp[-E_{\rm bh}/(k_{\rm B}T)]$, with $\nu_0$ = 17\ GHz and 
mean barrier height E$_{\rm bh}$ = 0.5\ K. The associated drop in $\nu_\mu$ at 
0.24\ K amounts to a factor of $10^3$.
Choosing a different value for $\Delta_{\rm HT}$/$\gamma_{\mu}$ would lead to 
different rates above 0.24\ K and to a different size for the drop.
For example, with  $\Delta_{\rm HT}$/$\gamma_{\mu}$ = 5.7\ mT (the value we
obtained for $\Delta_{\rm LT}$/$\gamma_{\mu}$ below 0.24\ K), the drop is 
$\simeq 10^2$ (this is also the minimum size of the drop allowed by the 
M\"ossbauer results) and with $\Delta_{\rm HT}/\gamma_{\mu}$ = 85.6\ mT, 
obtained by scaling from Tb$_2$Ti$_2$O$_7$ 
\cite{gardner}, it is $\simeq 10^4$. Combining the M\"ossbauer and the muon 
results thus indicates that the spin fluctuation rate undergoes a first 
order change by a factor of $10^2$ to $10^4$.
We mention that between 2 and 0.275~K, we find that $\lambda_z$ does not have
the usual Lorentzian dependence on the applied longitudinal field.
This anomalous behaviour will be discussed in a separate publication. 

The low temperature magnetic properties of Yb$_2$Ti$_2$O$_7$ are therefore 
unusual. In the short range correlated region 
from $\sim$ 2\,K to $\sim$ 0.24\,K, the spin fluctuation rates follow a 
thermal excitation law. This crystallographically ordered compound thus 
possesses barriers against spin reorientation.
On crossing the temperature of the specific heat $\lambda$-peak
\cite{blote}, no long range order appears, but there is a first order drop, of 
two to four orders of magnitude, in the fluctuation rates of the correlated 
moments. As $T \to 0$, the fluctuations persist at a temperature 
independent rate of $\simeq$ 1\ MHz, and they take place between directions 
which make an angle of $\sim 44^\circ$ relative to the 
local anisotropy axis (a [111] direction). The fluctuations of the moments 
thus conserve this angle and hence involve spin flips 
or spin spirals around a [111] axis. 

Above 0.24\ K, we speculate that the 
spin fluctuations could link more random directions.
The first order drop in the fluctuation rate would then be associated
with a change in the nature of the fluctuations such that below 0.24\,K, 
the system explores a subset of the states that are explored at higher 
temperatures.
The described transition, first order without the appearance of any long
range correlations and with a change of dynamics, evidences properties
which parallel aspects of the general gas-liquid transition or of the
specific water-ice transition.
To our knowledge, such a transition has not previously been observed in a
magnetic system.

When exotic low temperature properties have previously been observed in the 
rare earth pyrochlores, they have related to ions with Ising character 
(Tb$^{3+}$, Dy$^{3+}$, Ho$^{3+}$). 
The present results show that further novel behaviour 
is evidenced when the rare earth has a planar anisotropy. Although 
Yb$_2$Ti$_2$O$_7$ does not correspond to the spin-ice scenario (in addition to 
the planar anisotropy, there is, at most, only a small missing entropy), 
it shares with the spin-ice state the characteristic of short range 
correlated moments which continue to fluctuate between specific directions
as T $\to$ 0. Here however, the low temperature state is reached through a 
novel route: a first order dynamical transition which can be viewed as a 
change from a thermally excited regime to a low temperature quantum (or 
tunneling) regime.

\smallskip
We thank Isabelle Mirebeau and Collin Broholm for useful discussions and 
Nadine Genand-Riondet for technical assistance.

\end{document}